# Nanomechanical Spectroscopy of Ultrathin Silicon Nitride Suspended Membranes

S. S. Jugade, A. Aggarwal, A. K. Naik*

*Centre for Nano Science and Engineering (CeNSE), Indian Institute of Science (IISc) Bengaluru 560012, India*

*E-mail: anaik@iisc.ac.in

**Abstract**

Mechanical properties of a nanomechanical resonator have a significant impact on the performance of a resonant Nano-electromechanical system (NEMS) device. Here we study the mechanical properties of suspended membranes fabricated out of low-pressure chemical vapor deposited silicon nitride thin films. Doubly-clamped membranes of silicon nitride with thickness less than 50 nm and length varying from 5 µm to 60 µm were fabricated. The elastic modulus and stress in the suspended membranes were measured using Atomic Force Microscope (AFM)-based nanomechanical spectroscopy. The elastic modulus of the suspended membranes was found to be significantly higher than those of corresponding thin films on the substrate. A reduction in the net stress after the fabrication of suspended membrane was observed and is explained by estimating the contributions of thermal stress and intrinsic stress. We establish a mathematical model to calculate the normalized elastic modulus of a suspended membrane. Lastly, we study the capillary force-gradient between the $SiN_x$ suspended membrane-Si substrate that could collapse the suspended membrane.

**Keywords** – silicon nitride, thin films, suspended membranes, atomic force microscopy, elastic modulus, stress, capillary condensation, Nano-electromechanical systems.

Silicon nitride nanomechanical resonators have potential application as Nano-electromechanical systems (NEMS) sensors [1,2,3,4] due to their high-quality factor (in the order of few millions) and high sensitivity at room temperatures.[5,6,7,8] The fundamental natural frequency of a doubly-clamped membrane resonator depends on its elastic modulus and stress as given below.[9]

$$f_0 = 1.03 \sqrt{\frac{Et^2}{\rho L^4} + \frac{\sigma W t}{3.4 mL}} \tag{1}$$

where $E$ is the elastic modulus, $\sigma$ is the stress in the suspended membrane, $\rho$ is the density of the material, $m$ is the effective mass, $L, w$ and $t$ are the length, width and thickness of the suspended membrane respectively. Verbridge et al.[10] have shown that the addition of tensile stress to the silicon nitride NEMS resonators leads to higher quality factors which leads to better sensitivity. Furthermore, if the elastic modulus and stress in the suspended membrane are very low, then it is likely to collapse or undergo plastic deformation and fracture under a few nanonewtons of force.

For thin films grown on substrate, depending on the growth conditions, the mechanical properties can vary significantly from that of the bulk material.[11,12,13] The Low-Pressure Chemical Vapor Deposition (LPCVD) grown silicon nitride thin films studied in this work have a thickness less than 100 nm and are amorphous in nature.[14] Residual stress in thin films depend on the growth conditions such as pressure, temperature and gas flow ratio. The stress in an amorphous thin film can be separated into its intrinsic and thermal stress components.[15] Intrinsic stress results from defects such as voids, porosities, impurities and atomic diffusion. And thermal stress is the result of thermal expansion coefficient mismatch between the film and the substrate material.[16] High tensile stress in the deposited film leads to cracks in the film and subsequent failure of the fabricated suspended membrane. Hence, it becomes important to measure the elastic modulus and stress in silicon nitride thin films on substrate and suspended membranes fabricated out of them.

In this paper, first we describe the fabrication of silicon nitride suspended membranes. Then we explain the measurement techniques used to calculate the elastic modulus and stress in the SiN$_x$ suspended membranes and corresponding values on substrate thin films. The stress in the suspended membranes is studied based on relaxation of stress in the corresponding on substrate thin films. While the variation of elastic modulus with the thickness of the suspended membrane is analyzed using a model based on surface-energy considerations. In the end, the capillary force-gradient between the suspended membrane and bottom substrate is modeled based on the force-separation curves for the collapse of suspended membranes.

**Results and Discussion**

LPCVD process was used to deposit silicon nitride films of thickness 47 nm, 51 nm and 88 nm on a Si/SiO$_2$ substrate. The fabrication process of doubly-clamped membranes is illustrated in figure 1a. Figure 1b and 1c show a Scanning Electron Microscope (SEM) and Atomic Force Microscope (AFM) micrographs of a suspended membrane respectively. The suspended membrane shown in the figure is 5 μm in length and 500 nm in width. To prevent the collapse of suspended membranes due to the charging effect, we did not perform SEM imaging of the actual suspended membranes used for the measurements.

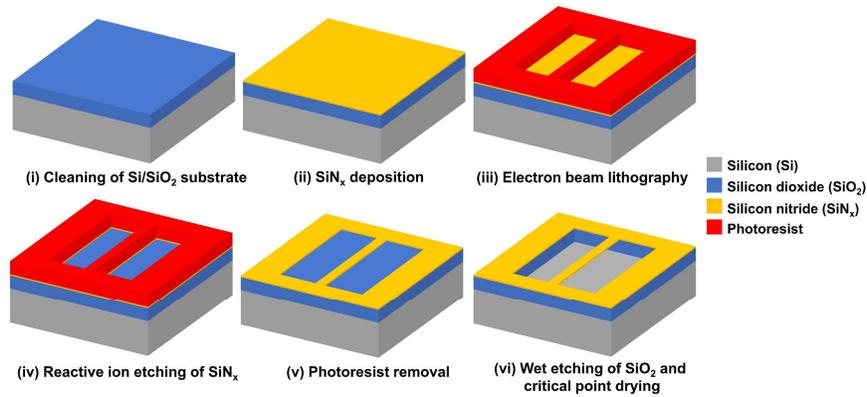

*a*

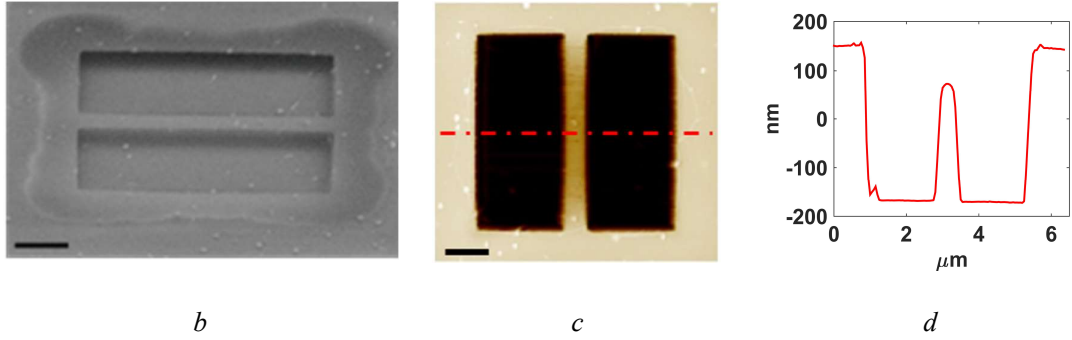

b           c           d

**Figure 1.** *(a) Process flow for the fabrication of a doubly-clamped silicon nitride membrane. (b), (c) scanning electron micrograph and AFM image of a 5 μm long and 500 nm wide suspended membrane respectively. (d) Topography of the section shown in figure 1c. Scale bar is 1 um for figure 1b and 1c.*

The elastic modulus and stress in the suspended membrane were measured using AFM based central indentation method.[9,17,18] Central indentation experiments were performed with AFM using the force-distance (F-d) spectroscopy.[19,20,21] The central indentation method is illustrated in figure 2a. Figure 2c show an AFM topography of a 5 μm long, 1.3 μm wide suspended membrane fabricated from a 51 nm thick silicon nitride film on substrate. A Force vs. Deformation curve (figure 2d) was obtained after central indentation on the suspended membrane. This curve was fitted using a force-deformation relation for a doubly clamped membrane subjected to central loading [22] as given in equation (2).

$$F = \left(16.23EW\left(\frac{t}{L}\right)^3 + 4.93\sigma\frac{Wt}{L}\right)\delta + 12.17E\frac{Wt}{L^3}\delta^3 \qquad (2)$$

where $E$ is the elastic modulus of the suspended membrane, $\sigma$ is the stress in the suspended membrane, $\delta$ is the deformation in the suspended membrane, $L, W$ and $t$ are the length, width and thickness of the suspended membrane. The coefficients of $\delta$ and $\delta^3$ terms in equation (2) include elastic modulus and stress. The elastic modulus and stress values were obtained after fitting the Force vs. Deformation curve to the experimentally observed data. For each thickness of the suspended membranes, the elastic modulus and stress measurements were performed on three different membranes of 5 µm length.

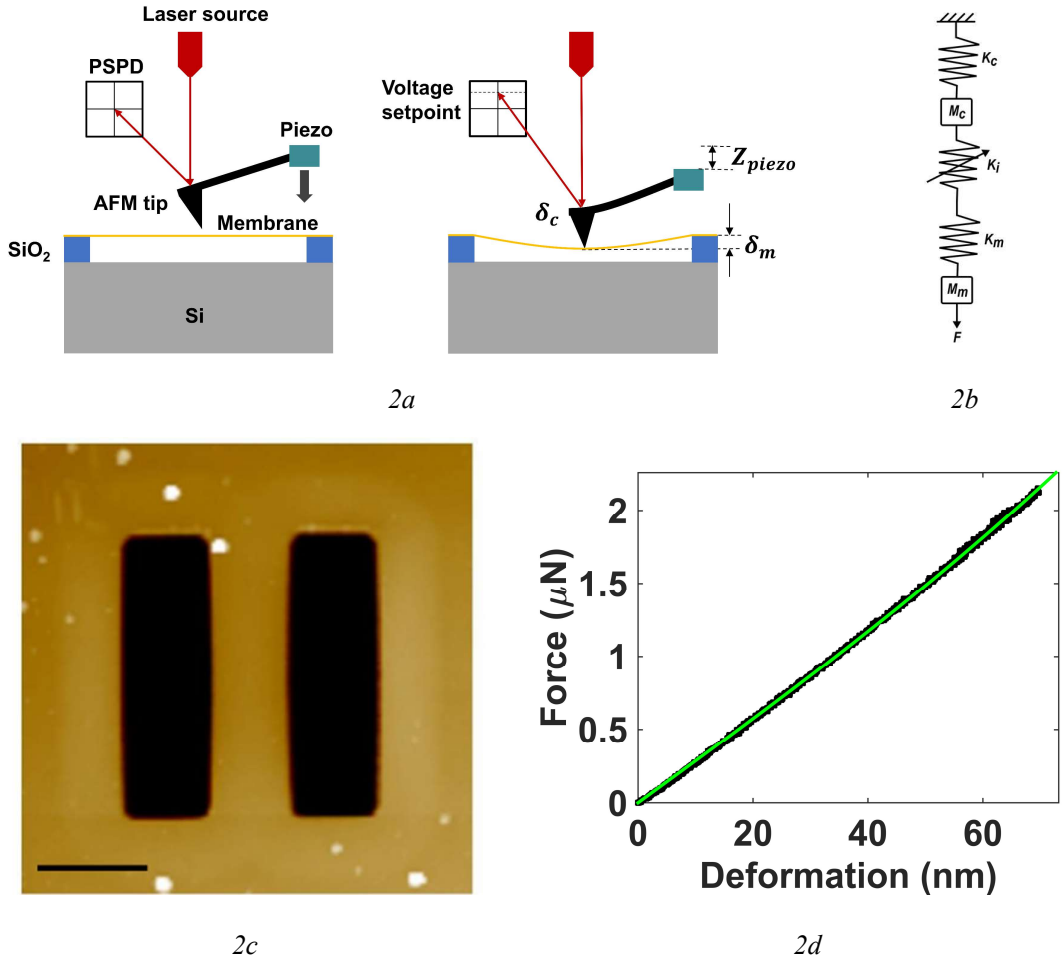

*Figure 2. (a) Schematic of the AFM based central indentation method. (b) Equivalent spring-mass model of the system - $K_c$, $K_i$, and $K_m$ are the stiffness of the AFM cantilever, tip-sample interaction, and membrane respectively. $K_i$ is a variable stiffness. F is the applied force, $M_c$ and $M_m$ are the effective mass of the cantilever and the membrane respectively. (c) AFM topography micrograph of 5 µm long and 1.3 µm wide membrane fabricated from a 51 nm thick silicon nitride film. Scale bar is 2 µm. (d) Fitted Force vs. Deformation curve obtained on the same membrane after central indentation experiment.*

For comparison with suspended membranes, the elastic modulus of the silicon nitride thin films on Si/SiO$_2$ substrate was measured. We use the nanoindentation [23] technique to perform these modulus measurements. The nanoindentation measurements were performed at four different locations on each of the thin films on substrate. To find the stress in these on substrate films, the change in curvature of Si/SiO$_2$ substrate due to deposition of SiN$_x$ film was measured using the beam deflection technique.[16] Stoney's equation was then used to calculate the stress in the SiN$_x$ film.[24] Due to a high deposition temperature of 750°C, thermal stress is developed in the film when it was cooled down to room temperature. The contribution of thermal stress [16] to the net stress can be calculated as given below.

$$\sigma_{th} = \frac{E_f(\alpha_s - \alpha_f)\Delta T}{(1-\vartheta_f)} \tag{3}$$

where $\sigma_{th}$ is the thermal stress in the film. $E_f$, $\alpha_f$ and $\vartheta_f$ are the elastic modulus, coefficient of thermal expansion and Poisson's ratio of the SiN$_x$ film material. $\alpha_s$ is the coefficient of thermal expansion of the substrate material. $\Delta T$ is the difference between the deposition temperature and room temperature. For calculations, the measured value of $E_f$ using nanoindentation method was used and average values of $\alpha_f = 3.2 \times 10^{-6}$/°C (for SiN$_x$) and $\alpha_s = 0.65 \times 10^{-6}$/°C (for SiO$_2$) were considered.[25,26] It is clear from equation (3), that the thermal stress will be larger in a film with higher elastic modulus than in a film with lower elastic modulus. Intrinsic stress in a film was calculated by subtracting the thermal stress from the net stress.

The elastic modulus, net stress, thermal stress, and intrinsic stress of the films on substrate are given in Table 1a. The elastic modulus of films of thickness 47 nm and 51 nm was very low i.e. about 10 % of the bulk modulus value of 270 GPa.[27] While the elastic modulus of 88 nm thick film was about 70% of the bulk value. The stress in each of the film was close to 1.5 GPa. The LPCVD process-parameters were same for all three films except for the gas flow ratio (NH$_3$: H$_2$Cl$_2$Si). For the 51 nm thick film, the gas flow ratio was 10:100 while for the other two films the ratio was 10:70. Increasing the gas flow ratio from 1:2 to 1:20 for nitrogen-rich LPCVD SiN$_x$ films has a negligible effect on the residual stress.[28,29] This explains why the net stress in all three films was almost equal. The contribution of thermal stress to the overall stress was negligible for the films of thickness 47 nm and 51 nm due to their low elastic moduli. While for the 88 nm thick film, thermal stress was about 30% of the total stress due to its higher elastic modulus.

The thickness of suspended membranes (Table 1b) were significantly lower than the thickness of corresponding film on substrate (Table 1a). This is the result of SiN$_x$ etch during the wet etching of sacrificial SiO$_2$ with BOE. This reduction in thickness after the fabrication of suspended membrane can be expressed by a ratio $\Delta t$ given by equation (4). Higher the value of $\Delta t$, lesser is the thickness of suspended membrane than the corresponding film on substrate. A high value of $\Delta t$ also indicates that that a large fraction of the highly-stressed region at the bottom of SiN$_x$ film on substrate is etched during the fabrication of suspended membrane. The values of $\Delta t$ are 0.83, 0.42, and 0.49 for the on-substrate films of thickness 47 nm, 51 nm and 88 nm respectively.

$$\Delta t = \frac{t_f - t_m}{t_f} \qquad (4)$$

***Table 1a.*** *Elastic modulus and stress in the on substrate SiN$_x$ thin films*

| Thickness of deposited SiN$_x$ film | Elastic Modulus of on substrate films (Nanoindentation method) | Net stress in on substrate films (Beam deflection method) | Thermal stress in on substrate films (Calculated using equation 3) | Intrinsic stress in on substrate films |
|---|---|---|---|---|
| $t_f$ (nm) | $E_f$ (GPa) | $\sigma_f$ (GPa) | $\sigma_{th}$ (GPa) | $\sigma_i$ (GPa) |
| 47 | 20.54 ± 1.985 | 1.432 ± 0.114 | 0.054 | 1.378 |
| 51 | 28.52 ± 0.528 | 1.530 ± 0.190 | 0.075 | 1.455 |
| 88 | 184.5 ± 13.27 | 1.627 ± 0.259 | 0.487 | 1.140 |

***Table 1b.*** *Elastic modulus and stress of the 5 μm long SiN$_x$ suspended membranes measured using AFM*

| Thickness of deposited SiN$_x$ film on substrate $t_f$ (nm) | Thickness of SiN$_x$ suspended membrane $t_m$ (nm) | Elastic Modulus of suspended membrane $E_m$ (GPa) | Stress in suspended membrane $\sigma_m$ (MPa) | Spring constant of suspended membrane $k_m$ (N/m) |
|---|---|---|---|---|
| 47 | 8  | 104.1 ± 1.2  | 241.7 ± 1.5 | 0.233 |
| 51 | 30 | 222.6 ± 9.1  | 861.8 ± 2.7 | 1.845 |
| 88 | 45 | 204.1 ± 16.9 | 552.7 ± 4.5 | 3.260 |

It can be observed from Table 1a and 1b that stress values in the suspended membranes were significantly lower than those in the corresponding films on substrate. This reduction in stress after the fabrication of suspended membrane was due to relaxation of thermal and intrinsic stress. The thermal stress relaxes because the suspended membrane was no longer attached to the substrate except at its clamped ends. The relaxation of thermal stress was negligible in suspended membranes of thickness 8 nm and 30 nm, while a 487 MPa of thermal stress was relaxed in case of 45 nm thick suspended membrane. The relaxation of intrinsic stress depends on the value of $\Delta t$ for the particular suspended membrane. The value of $\Delta t$ for the suspended membrane of thickness 8 nm was 0.81, highest amongst all the three suspended membranes. This implies that most of the highly stressed regions of the 47 nm thick SiN$_x$ film near the substrate were etched after fabrication of the suspended membrane. Hence, there was a drastic reduction in stress from 1.432 GPa (for 47 nm thick on substrate film) to 242 MPa (for 8 nm thick suspended membrane). The thermal stress in the 88 nm thick film was about six times higher than that in the 51 nm thick film (Table 1a). Hence, the net reduction in stress after the fabrication of membranes is also higher (Table 1b). In fact, simple calculations reveal that relaxation of intrinsic stress in case of the suspended membranes of thickness 30 nm (51nm thick film on substrate) and 45 nm (88nm thick film on substrate) was 593 MPa and 587 MPa respectively. So, the relaxation of intrinsic stress was approximately equal. This is also reflected in the close values of 0.42 and 0.49 for $\Delta t$ of these two suspended membranes.

From Table 1a and 1b, it can also be observed that the elastic modulus of the suspended membranes were significantly higher than those of the corresponding films on the substrate. The 30 nm thick suspended membrane had the highest elastic modulus of 222 GPa. We expect that the calculated elastic moduli of the suspended membranes are slightly overestimated. The two main reasons for this are the surface effects and offset of the tip from the center of the suspended membrane during the indentation. In AFM based indentation methods, repulsive interaction forces between the tip and surface of the specimen leads to a value of elastic modulus higher than the one obtained from nanoindentation techniques.[30] Thermal drift in the piezo and inclined motion of tip creates an offset of few tens of nanometers from the center of the suspended membrane. The motion of the tip is not in a direction perpendicular to the surface of suspended membrane owing to the tip-sample repulsive forces. These effects lead to a higher value of calculated elastic modulus. The coefficient of linear deformation term ($\delta$) in equation (2) gives the spring constant of the suspended membrane. The 8 nm thick suspended membrane was around 8 times softer than the 30 nm thick suspended membrane. While the suspended membranes of thickness 30 nm and 45 nm had relatively same values of spring constant. We also observed a hysteretic mechanical behavior of 8 nm thick membranes for higher applied peak loads (section 3 of supplementary information).

The trend in the values of elastic modulus is difficult to understand as it is not monotonic. Hence, the elastic modulus of suspended membranes was further investigated using and approach that is similar to the one followed by Fedorchenko et al.[31] A suspended membrane has both the top and bottom surfaces free. So, for surface energy calculations it can be considered as a thin film with both surfaces free. For such a film, the limit thickness $t_0$ below which it cannot be considered as a bulk material [31] is given in equation (5).

$$t_0 = \frac{(1-\vartheta_f)}{E_{bulk}\varepsilon_0{}^2} \tag{5}$$

where $\vartheta_f$ is the Poisson's ratio of the film = 0.3 (assumed), $E_{bulk}$ is the bulk elastic modulus of the film material, $\varepsilon_0$ is the film strain. Table 2 lists the calculated values of $t_0$ and $\varepsilon_0$ for the suspended membranes. The strain in all the suspended membranes was not equal. Hence, a normalized thickness of a suspended membrane was defined as the ratio of the thickness of suspended membrane and the limit thickness. Similarly, the normalized elastic modulus of a suspended membrane was also defined.

**Table 2.** *Limit thickness ($t_0$) and strain ($\varepsilon_0$) values for the suspended membranes*

| Thickness of the suspended membrane $t_m$ (nm) | Strain in the suspended membrane $\varepsilon_0$ ($\times 10^{-3}$) | Limit thickness $t_0$ (nm) | Normalized thickness of the suspended membrane $N_t = \dfrac{t_m}{t_0}$ | Normalized elastic modulus of the suspended membrane $N_E = \dfrac{E_m}{E_{bulk}}$ |
|---|---|---|---|---|
| 8 | 2.7 | 347.47 | 0.023 | 0.328 |
| 30 | 4.6 | 124.97 | 0.240 | 0.700 |
| 45 | 3.2 | 255.43 | 0.176 | 0.642 |

The normalized thickness of suspended membranes clearly explains the calculated elastic modulus data. The 30 nm thick suspended membrane has the highest normalized thickness hence it has the highest elastic modulus among all the three suspended membranes. The variation in the normalized elastic modulus of suspended membrane vs. the normalized thickness of the suspended membrane can be expressed using equation (6).

$$N_E = (N_t)^p \tag{6}$$

Curve fitting yields a value of $0.27 \pm 0.06$ for $p$. The fitted curve is shown in figure 3. For any value of strain and thickness of the suspended membrane, equation (6) can be used to determine the elastic modulus of the suspended membrane.

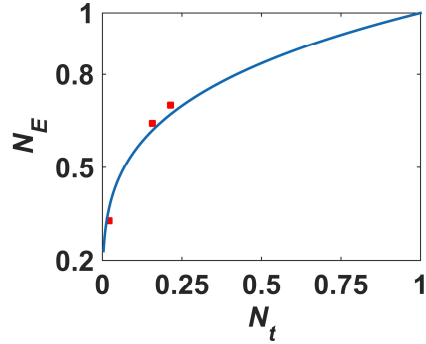

**Figure 3.** *Normalized elastic modulus vs. Normalized thickness of the suspended membrane.*

The suspended membranes of thickness 30 nm collapsed to the bottom substrate under large applied loads at the center of the suspended membranes. Due to condensation of a water droplet, a liquid bridge (capillary) is formed between the surface of a suspended membrane and the bottom substrate when they are very close to each other.[32,33] This capillary adhesion force between the two surfaces collapses the suspended membrane. The pull-in of the suspended membranes is illustrated in the force-separation curves in figure 4a, 4b and 4c. The experimental parameters of the pull-in measurements are listed in Table S2 of supplementary information. The point of the collapse of a suspended membrane is indicated by a sudden drop in force in the corresponding force-separation curve. Such an abrupt drop in force was due to the relaxation of the cantilever as the suspended membrane moves down. During the pull-in of the suspended membrane, the Z position of piezo did not change as the pull-in process took less than 1 milliseconds time. The pull-in process time ($t_c$) is three orders of magnitude greater than the inverse of the fundamental resonance frequency of the suspended membranes ($1/f_0$). Hence, the dynamics of pull-in of the suspended membranes is governed by viscous damping.

The net relaxation of the AFM cantilever in the region R (figure 4d, 4e, and 4f) is equal to the gap between the suspended membrane and substrate at the start of the pull-in process. For the suspended membranes of length 20 µm and 40 µm, the AFM cantilever did not relax completely after the pull-in of the suspended membrane was over (stage (iv) in figure 5a). While for the 60 µm long suspended membrane, the cantilever relaxed completely before the pull-in process was completed and then lost contact with the suspended membrane. As a result of the small gap, capillary condensation occurred between the AFM cantilever and the suspended membrane that is now attached to the substrate (stage (iv) in figure 5b). This lead to the snap-in of the AFM cantilever to the surface of suspended membrane (stage (v) in figure 5b).

In all the three suspended membranes, the piezo moves down in Z direction and the tip presses down on the hard $SiN_x$ on Si substrate after the pull-in of suspended membrane is over. During this process, only the AFM cantilever is deflected and that is why an infinite slope is observed in the force-separation curves. The cantilever continues to deflect until the force setpoint is reached (stage (v) and (vi) in figure 5a and 5b respectively). After this, the piezo retracts and cantilever pulls-off from the $SiN_x$ surface by overcoming the adhesion force.

A suspended membrane collapses when the sum of capillary force gradient and stiffness of suspended membrane becomes zero as given in equation (8). The stiffness of suspended membrane was calculated as negative of the slope of Force vs. Tip-sample separation curve in the region just before the collapse of the suspended membrane. The obtained Capillary force gradient vs. gap data was fitted using a power relation (equation 9) as shown in figure 6a.

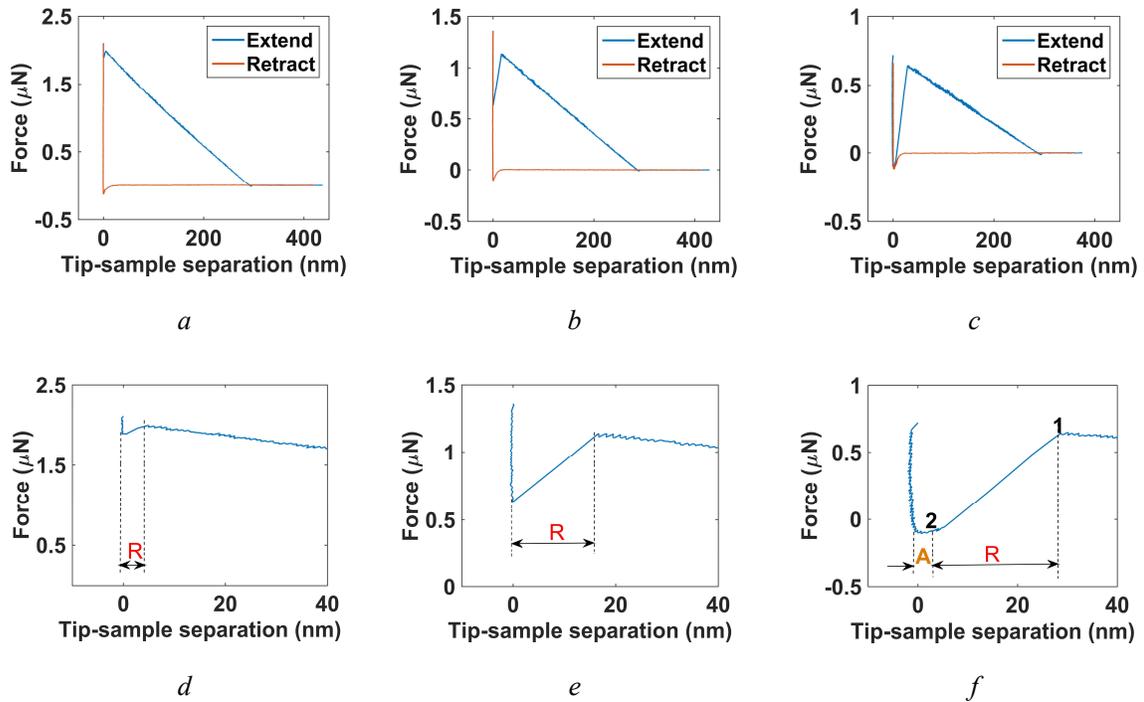

*Figure 4.* Force vs. Tip-sample separation curves indicating the collapse of 51 nm thick suspended membrane. *(a)* For 20 μm long suspended membrane. *(b)* For 40 μm long suspended membrane. *(c)* For 60 μm long suspended membrane. *(e), (f), (g)* Zoomed region of the Force vs. Tip-sample separation extend curve corresponding to the pull-in of suspended membranes of length 20 μm, 40 μm and 60 μm respectively. The slope of the curve in region R is equal to the spring constant of the cantilever.

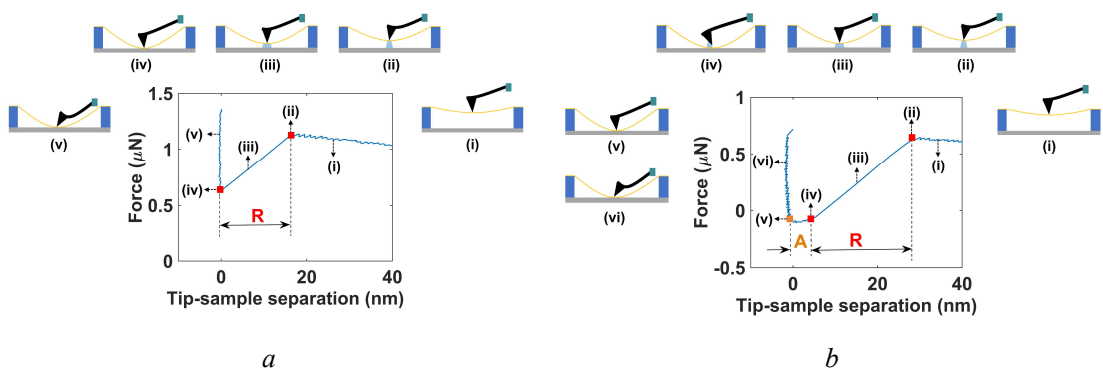

*Figure 5. (a)* Pull-in process illustrated for suspended membranes of length 20 μm and 40 μm based on the force-separation curve for 40 μm long suspended membrane. *(b)* Pull-in process illustrated for 60 long suspended membrane based on its force-separation curve.

$$\frac{\partial F}{\partial h} + K_m = 0 \qquad (8)$$

$$\frac{\partial F}{\partial h} = -138.6\, h^{-2.03 \pm 0.001} \qquad (9)$$

where, $\frac{\partial F}{\partial h}$ is the capillary adhesion force-gradient, $K_m$ is the stiffness of the suspended membrane and $h$ is the gap between the SiN$_x$ suspended membrane and the bottom Si substrate.

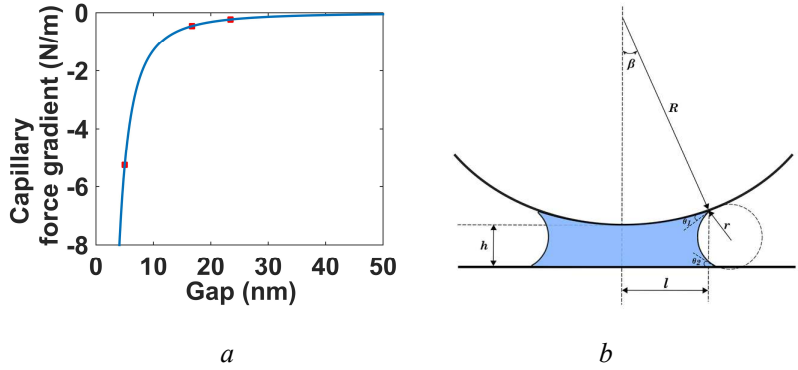

*Figure 6. (a) Capillary force gradient vs. Gap between the suspended membrane and the bottom substrate. (b) Schematic of a model for the capillary force between a sphere and a flat plate.*

Integration of equation (9) shows that the capillary force varies as $\frac{1}{h^{1.03}}$ with the distance between the suspended membrane and surface of bottom substrate. Consider a section parallel to the width of a suspended membrane and passing through the center of a suspended membrane. For this cross-section, the central portion of the suspended membrane has a finite radius of curvature because the width of the suspended membrane ($\sim 1\,\mu m$) is comparable to the deflection ($\sim 300\,nm$). Therefore, an expression for the capillary force between the suspended membrane and bottom substrate can be determined analytically using a sphere and flat plate model shown in figure 6b.[34] Comparing the analytical expression (for small $\beta$) and experimental fit gives $R \sin \beta = 12.293$, where R is in nm. Thus, a known value of either the radius of curvature of suspended membrane ($R$) or the filling angle ($\beta$) would be sufficient to determine the variation of capillary force with the gap between the suspended membrane and bottom substrate.

**Conclusions**

In summary, we fabricated silicon nitride suspended membranes of thickness 8 nm, 30 nm, and 45 nm from the LPCVD high-stress $SiN_x$ films of thickness 47 nm, 51 nm, and 88 nm respectively. Using AFM based indentation method, we studied the variation in elastic modulus and stress of a 5 μm length suspended membrane for three different thickness. Thermal stress in the films is relaxed after the fabrication of suspended membrane but the relaxation of intrinsic stress depends on the fractional reduction in thickness ($\Delta t$) after the fabrication of suspended membrane. Higher the value of $\Delta t$, higher will be the relaxation of intrinsic stress after fabrication of the suspended membrane. The obtained values of elastic modulus in these suspended membranes are higher than the actual value due to surface effects and offset between the tip and center of the suspended membrane. A model is developed to determine the variation of normalized elastic modulus with the normalized thickness of the suspended membrane. This understanding of the variation of elastic modulus and stress in the suspended membranes is important for developing more efficient NEMS resonators.

Finally, the collapse of 30 nm thick suspended membranes under large deformation is explained based on the capillary condensation and force-separation curves. And it is found that the capillary force-gradient varies inversely as square of the gap between suspended membrane and bottom substrate. This result can be particularly helpful in the design and fabrication of suspended membrane-based NEMS devices so as to prevent failure due to condensation induced capillary adhesive forces

**Methods**

**Fabrication of SiN$_x$ suspended membranes**

The steps involved in the fabrication process were: (i) RCA cleaning of the Si/SiO$_2$ substrate. (ii) Deposition of LPCVD grown high-stress silicon nitride thin film on the substrate. The process parameters are listed in Table S1 of supplementary information. (iii) Spin-coating of PMMA A4 photoresist, followed by patterning of suspended membrane-features for different dimensions using Electron beam lithography and then the development of photoresist using MIBK: IPA (1:3) developer. (iv) Reactive-ion-etching of silicon nitride from the windows opened in the lithography step using SF$_6$ gas. (v) Removal of PMMA using oxygen plasma. (vi) Wet etching of sacrificial silicon dioxide using 13:2 buffered oxide etch (BOE/BHF), followed by Critical Point Drying (CPD) to release the suspended membranes.

**AFM characterization of SiN$_x$ suspended membranes**

The topography of the suspended membranes was acquired in AFM by scanning with a constant force of 10-40 nN in Peak Force Quantitative Nanomechanical Mapping (PFQNM) [35] mode in Bruker Dimension Icon AFM. The length and width of the suspended membrane are determined from the topography micrograph. The thickness of a suspended membrane is measured after acquiring the force-distance curves on it. A large force of about 10 µN is applied to the suspended membrane during the AFM scan. This force was sufficient to break the suspended membrane. The step height is then measured in the region where the fractured suspended membrane is in contact with the bottom Si substrate. The thickness of a suspended membrane was found to be uniform in a large area of fractured portions of the suspended membrane in contact with the bottom substrate.

**Measurement of thickness of SiN$_x$ thin films**

The thickness of SiN$_x$ films and SiO$_2$ layer was measured using Variable Angle Spectroscopic Ellipsometer (VASE) [36] J.A. Woollam Co. M2000U. The measurement is performed at an incident angle of 65°, 70°, and 75°. The thickness of the SiO$_2$ layer is 300 nm.

**Measurement of elastic modulus and stress in on substrate SiN$_x$ thin films**

The elastic moduli of the silicon nitride thin films were measured using nanoindentation technique. Hysitron TI 950 Triboindenter with a Hysitron spherical diamond tip (radius 1 μm) is used to indent the sample in load control mode. Loading and unloading Force vs. Displacement curves are acquired in a load control of 100 μN, 180 μN and 500 μN for the films of thickness 47 nm, 51 nm, and 88 nm respectively. The elastic modulus is obtained by fitting the unloading curve using the Oliver-Pharr model.[37,38] To measure the stress of SiN$_x$ films, the beam deflection technique was employed in a laser-based kSA Multi-beam Optical Sensor (MOS) system. The change in the curvature of the substrate is converted to stress in the film using Stoney's equation.

**Acknowledgments**

We acknowledge funding support from DST SERB through grant number EMR/2016/006479 and from LAM Research through 2018 Unlock ideas campaign. We also acknowledge funding support from MHRD, MeitY and DST Nano Mission through NNetRA. We thank Prosenjit Sen for helpful technical discussions.

# Nanomechanical Spectroscopy of Ultrathin Silicon Nitride Suspended Membranes

S. S. Jugade, A. Aggarwal, A. K. Naik*

*Centre for Nano Science and Engineering (CeNSE), Indian Institute of Science (IISc)*

*Bengaluru 560012, India*

## Supplementary Information

**1. Low-Pressure Chemical Vapour Deposition (LPCVD) process parameters**

The silicon nitride thin films were deposited on a Si/SiO$_2$ substrate by Low-Pressure Chemical Vapour Deposition (LPCVD) process. The important process-parameters are listed below.

*Table S1. LPCVD process-parameters*

| Thickness of SiN$_x$ film (nm) | Gas-flow ratio (Dichlorosilane: ammonia) (sccm:sccm) | Deposition pressure (mTorr) | Deposition temperature (°C) |
|---|---|---|---|
| 47 | 10:70 | 300 | 750 |
| 51 | 10:100 | 300 | 750 |
| 88 | 10:70 | 300 | 750 |

**2. Central indentation method**

Figure 2a illustrates the measurement setup for the central indentation experiment. Initially, the AFM tip is located above the center of the suspended membrane at a distance where there was no interaction between the atoms of the tip and suspended membrane. Then the piezo moves down in Z direction resulting into deformation of the suspended membrane and deflection of the AFM cantilever. The piezo extends down until a voltage setpoint is reached on the position sensitive photodiode (PSPD). Then it retracts to its original position. Thus,

Voltage vs. Z-piezo curve was obtained for extending and retract motion at a rate of 1 Hz. The velocity of piezo during extension and retraction motion is 1 μm/sec.

The Voltage vs. Z-piezo curves are obtained after central indentation on the suspended membrane shown in the AFM image of fig. 2c. Based on the deflection sensitivity and spring constant calibration values [1] these curves were converted to Force vs. Z-piezo curves as shown in figure S1a. The Force vs. Z-piezo curves were then converted to Force vs. Separation curves (figure S1b) using Eq. (SE1).[2] The applied force and corresponding deformation were obtained by calculating the separation value for the jump to contact point when the tip approaches the sample (figure S1c). Thus, a Force vs. Deformation curve was obtained as shown in figure 2d.

$$Z_{piezo} = \delta_m + \delta_c \qquad (SE1)$$

where, $Z_{piezo}$ is the displacement of the piezo in the Z direction, $\delta_m$ is the deformation of the suspended membrane and $\delta_c$ is the deflection of the cantilever.

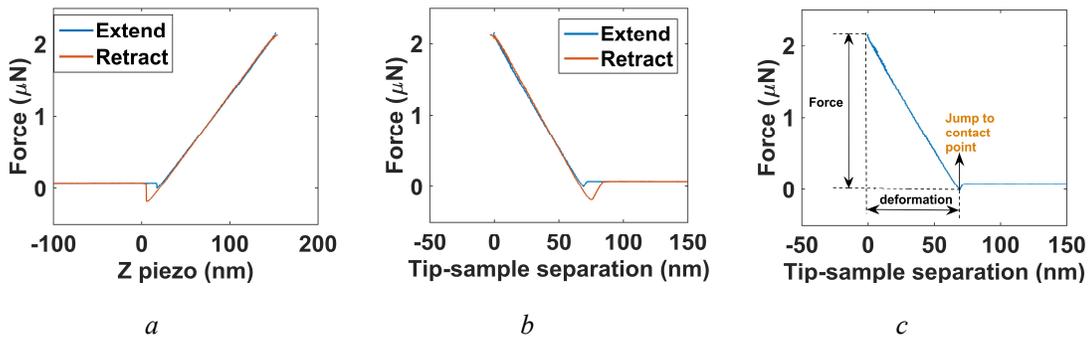

*Figure S1. (a), (b), (c) Force vs. Z piezo, Force vs. Tip-sample Separation curves and Extend Force vs. Tip-sample separation curve obtained after central indentation experiment on the suspended membrane in figure 2c in the manuscript respectively.*

The spring constant of the selected AFM cantilever was such that there was significant deformation of the suspended membrane with a detectable voltage difference on PSPD. The spring constant of the AFM cantilever used for the suspended membranes fabricated from 51 nm and 88 nm thick films was 31 N/m. For the relatively softer suspended membranes fabricated from 47 nm thick film, a cantilever of spring constant 17 N/m was used. The spring constant was calibrated using the thermal tuning.[3] The indentation depth on $SiN_x$ films using the above cantilevers was less than 0.5 nm. Hence, there was no pressing effect of the tip [4] during the indentation of the suspended membranes. The tip radius was very small (< 40 nm) compared to the suspended membrane length (> 5 μm) and width (> 1 μm). This ensured that the elastic modulus and stress values were not affected by the tip radius.

## 3. Hysteresis in the mechanical behavior of 8 nm thick suspended membranes

Fig. S2 shows the hysteresis in force-separation curves for 8 nm thick suspended membranes.

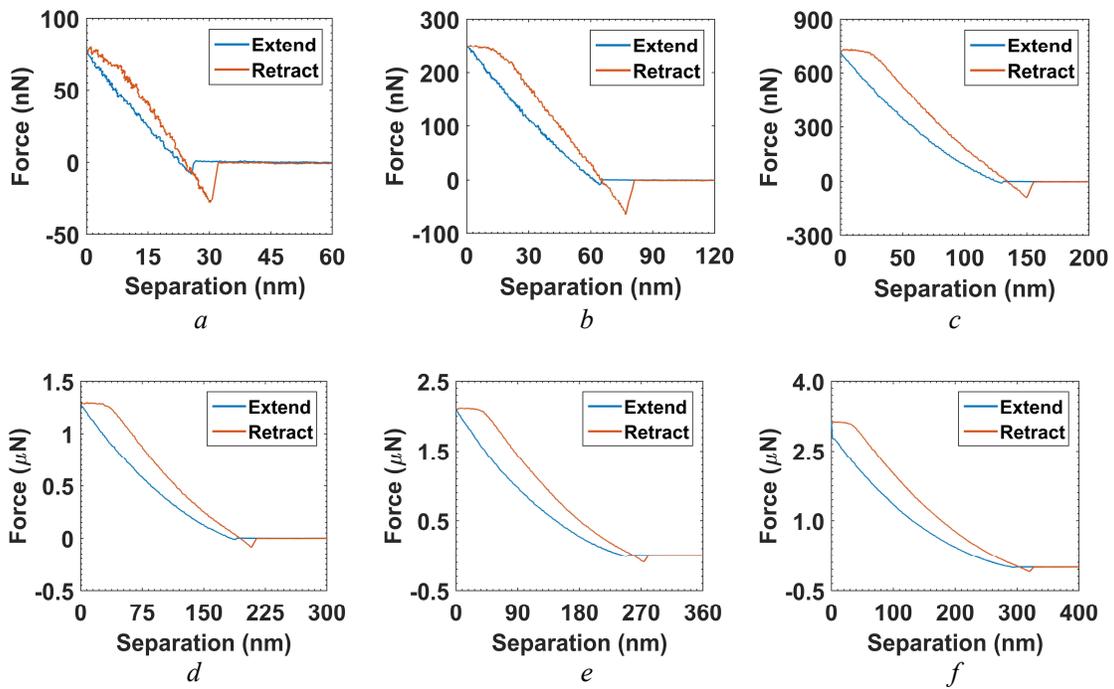

***Figure S2.*** *Hysteresis in force-separation curves for increasing peak force values in case of an 8 nm thick suspended membrane.*

The hysteresis is not significant for lower peak forces (fig. S2 a) but as the peak force increases beyond 100 nN, hysteretic behavior of suspended membrane increases. The extend and retract curves do not overlap in the repulsive region because, at the start of the extend cycle, the slope of the force-separation curve is zero (fig. S2 b-f). This zero slope indicates that the deflection of suspended membrane reduces at a constant force. This hysteretic mechanical behavior might indicate the viscoelastic nature of the 8 nm thick suspended membranes.

## 4. Experimental parameters for the pull-in of suspended membranes

The experimental parameters for the pull-in of suspended membranes is caluclated from the force-separation curves and is listed in the table below.

*Table S2. Experimental parameters during pull-in measurements*

| Parameters | Length of suspended membrane | | |
| --- | --- | --- | --- |
|  | 20 μm | 40 μm | 60 μm |
| Force on the cantilever at the start of the pull-in process, $F_{cs}$ (nN) | 1994 | 1140 | 643.9 |
| Gap between the suspended membrane and substrate at the start of the pull-in process, $h$ (nm) | 5.014 | 16.72 | 23.438 |
| Force on the cantilever at the end of the pull-in process, $F_{ce}$ (nN) | 1888 | 628.6 | -101.9 |
| Capillary force gradient, $\frac{\partial F}{\partial h}$ (N/m) | -5.255 | -0.4559 | -0.2296 |
| Pull-in process time, $t_c$ (μs) | 237 | 475 | 950 |
| (Fundamental resonance frequency)$^{-1}$, $1/f_0$ (μs) | 0.0526 | 0.0982 | 0.138 |

## 5. Model for capillary force-gradient

An analytical model for the capillary force between a sphere and flat plate [5] is used to model the capillary force between the suspended membrane and bottom Si substrate. The symbols in figure 6b are as follows: $R$ is the radius of sphere which is equal to the radius of curvature at center of the suspended membrane, $\beta$ is the filling angle, $r$ and l are the radii of curvature of meniscus, $\theta_1$ and $\theta_2$ are the contact angles of the liquid to the sphere and flat plate respectively and $h$ is the gap between the sphere and flat plate. Capillary force between the sphere and flat plate is given in equation (SE2). The expressions for the radii of curvature of meniscus are given in equation (SE3) and (SE4) respectively. It is important to mention that the radii of curvature and thus the capillary force are calculated at the three-phase contact line or at the filling angle $\beta$.

$$F = \pi \gamma R \sin \beta \left[ 2 \sin(\theta_1 + \beta) + R \sin \beta \left( \frac{1}{r} - \frac{1}{l} \right) \right] \quad \text{(SE2)}$$

$$r = \frac{R(1 - \cos\beta) + h}{\cos(\theta_1 + \beta) + \cos \theta_2} \quad \text{(SE3)}$$

$$l = R \sin \beta - r[1 - \sin(\theta_1 + \beta)] \quad \text{(SE4)}$$

The surfaces of $SiN_x$ membrane and bottom Si substrate should have a native oxide layer on it. Therefore, both the surfaces are hydrophilic and hence the contact angles of the meniscus should be zero. The assumption that $\beta$ is small is due to the fact that only a small central portion of the width of the membrane is involved in forming the meniscus. For $\theta_1 = \theta_2 = 0°$ and small filling angle $\beta$, the radii of curvature are $r = \frac{h}{2}$ and $l = \frac{-h}{2}$ and the capillary force is given by equation (SE5). The partial derivative of the capillary force w.r.t. the gap between the sphere and flat plate gives the capillary force-gradient given in equation (SE6).

$$F = 2\pi \gamma R (\sin \beta)^2 \left( 1 + \frac{2R}{h} \right) \quad \text{(SE5)}$$

$$\frac{\partial F}{\partial h} = \frac{-4\pi\gamma R^2 (\sin\beta)^2}{h^2} \tag{SE6}$$

It can be observed that the capillary force-gradient varies inversely as the square of $h$. Taking the value of surface of tension of water ($\gamma$) as 73 mN/m and then comparing equation (SE6) and equation (9) in the manuscript gives $R \sin\beta = 12.293$, where $R$ is in nm. $R$ changes with the gap between the suspended membrane and substrate. This implies that the filling angle $\beta$ also varies with the gap. Thus, knowing the value of $R$ or $\beta$ at a specific value of $h$ will be sufficient to completely determine the capillary force using equation (SE5).